\documentstyle[11pt,epsf]{article}
\def\lord{$ \raisebox{-.3ex}{$\stackrel{<}{_{\sim}}$} $}

\def\be{\begin{eqnarray}}
\def\ee{\end{eqnarray}}
\def\dis{\displaystyle}
\begin{document}

{\Large\bf
Strangeness in Stellar Matter 
}\\[3mm] 
\def\rightmark{Strangeness in Stellar Matter}\def\leftmark{Prakash, et al.}
\hspace*{6.327mm}\begin{minipage}[t]{12.0213cm}{\large\lineskip .75em
M. Prakash$^1$, S. Reddy$^1$, J. M. Lattimer$^2$ and P. J. Ellis$^3$ 
}\\[2.812mm] 
\hspace*{-8pt}$^1$ Physics Department, \\
State University of New York at Stony Brook, \\
Stony Brook, New York-11794, USA \\[0.2ex]
\hspace*{-8pt}$^2$ Earth and Space Sciences Department, \\
State University of New York at Stony Brook, \\
Stony Brook, New York-11794, USA \\[0.2ex]
\hspace*{-8pt}$^3$ School of Physics and Astronomy, \\
University of Minnesota, \\
Minneapolis, MN-55455, USA
\\[4.218mm]{\it
}\\[5.624mm]\noindent
{\bf Abstract.} 
A protoneutron star is formed immediately after the gravitational collapse of
the core of a massive star.  At birth, the hot and high  density matter in such
a star contains a large number of neutrinos  trapped during collapse.  
Trapped neutrinos generally inhibit the presence of exotic matter --  hyperons,
a kaon condensate,  or quarks.  However, as the neutrinos diffuse out in about
10-15 s, the threshold for the appearance of strangeness is reduced; hence, the
composition and the structure of the star can change significantly.  The effect
of  exotic,  negatively-charged, strangeness-bearing components is always to
soften the  equation of state, and the possibility exists that the star
collapses to a black hole at this time.  This could explain  why no neutron
star has yet been seen in the remnant of supernova  SN1987A, even though one
certainly existed when neutrinos were detected on Feb. 23, 1987.  With new
generation neutrino detectors it is feasible to test different  theoretical
scenarios observationally.              
\end{minipage}
 
\section{Introduction}

\noindent The possible existence of matter with strangeness to baryon ratio, 
$|S|/B$, of order unity has received a great deal of attention recently.  The
most likely site in which strangeness-bearing matter may exist is thought to
be the interior regions of a neutron star  (see Fig. 1).  The physical state
and internal constitution of neutron stars chiefly depends on the nature of the
strong interactions.  Although the composition and the equation of state (EOS)
of neutron star matter are not yet known with certainty, QCD based effective
Lagrangians have opened up intriguing possibilities.   Strangeness may occur in
the form of baryons, notably the $\Lambda$ and $\Sigma^-$ hyperons, or as a
Bose condensate, such as a $K^-$ meson condensate, or in the form of strange
quarks in a mixed phase of hadrons and quarks. {\em All these possibilities 
involve negatively charged, strange matter}, which, if present in dense matter,
 results in important consequences for neutron stars. 
For example, the appearance of strangeness-bearing components
results in protoneutron (newly born) stars  having larger maximum masses than
catalyzed  (older, neutrino-free) neutron stars, a reversal from ordinary
nucleons-only matter. This permits the existence of metastable protoneutron
stars that could collapse to black holes during their
deleptonization.  This  scenario
is outlined in Sec. 4 (for further details see Refs. \cite{prep,comments}).

\section{Some Observed Neutron Star Properties}

\noindent To date, the observations relevant to the understanding of neutron
star composition, structure, and evolution are mainly those of masses and
surface temperatures. 

\subsection{Masses}

Observed neutron star masses are shown in Fig. 2. The smallest range that
is consistent with all of the data has an upper limit of $1.44M_\odot$ from
PSR1913+16 and a lower limit of $1.36M_\odot$ from the precisely measured total
mass of PSR 2127+11C and its companion. The upper limit, $M=1.44M_\odot$,
provides constraints on the neutron star EOS.  
A conservative estimate of this upper limit is shown by
the dashed line at $M=1.5M_\odot$. Any neutron
star EOS has to support a maximum mass of at least this value.

The fact that all the measured neutron star masses consistently lie within a
narrow range of $1.4M_\odot$ has raised many intriguing issues.  One possible
explanation stems from evolutionary arguments.  Since neutron stars are formed
in the gravitational core collapse of massive stars, their masses may depend on
the structure of the progenitor star.  The cores of stars which evolve into
neutron stars have precollapse masses of about $1.4M_\odot$,  which
introduces a natural mass scale for possible neutron star masses.   The final
neutron star mass may, however, depend on the amount of accretion at times 
subsequent to a neutron star's birth. 
Thus, rigorous arguments for the
happenstance of observations are not yet available.  Later we will return to
the question of whether or not it is the nature of strong interactions which
restricts the stars to this mass range.  

\subsection{ Thermal emissions and surface temperatures }
    
Neutron stars are born with interior temperatures of order 20-50 MeV, but cool
via neutrino emission to temperatures of less than 1 MeV within
minutes~\cite{birth}. The subsequent cooling consists of two phases: a
neutrino-dominated cooling epoch 
followed by a photon-dominated cooling epoch.
The temperature and luminosity of this thermal radiation is controlled by the
evolution of the  interior temperature of the star. In the standard scenario,
in which the cooling occurs by the so-called modified Urca processes, $n + n
\rightarrow n + p + e^- + {\overline \nu}_e$ and  $n + p + e^- \rightarrow n +
n + \nu_e$,  the interior cooling is slow enough that the surface temperatures
of neutron stars remain above $10^6$ K for about $10^5$ yr, and they are
potentially observable for this length of time in the X-ray or UV bands. 
Recently, it has been pointed out~\cite{LPPH} that %reordered here down
a rapid cooling may occur via the direct 
Urca processes, $n \rightarrow p + e^- + {\overline \nu}_e$
and  $p + e^- \rightarrow n + \nu_e$,  and similar processes involving muons 
if the proton fraction 
in super-nuclear matter reaches values in excess of 
about $10\%$.  This occurs when exotic matter, such as a pion or a 
kaon condensate or quark matter, is present. Similar direct Urca processes 
involving strangeness-bearing hyperons also lead to rapid 
cooling~\cite{PPLP}.  In this case the core cools so rapidly that a
temperature inversion develops. The size of the cooler interior grows as the
energy from  the hot crust is conducted to the core. After about 1 to 100 yr,
depending on the stars's structure, this cooling wave reaches the surface, and
the surface temperature plummets to very low values. 
Thus, observations of 
surface temperatures hold the potential of revealing the internal 
constitution of neutron stars~\cite{Peth,Prak}.

Through recent advances in imaging X-ray telescopes, it has been possible to
detect X-rays from some 14 rotation-powered neutron stars (see, for example,
Ref. \cite{Ogel}).  Figure 3 shows the
inferred surface temperatures of some  X-ray emitters. Of these, four appear to
have the signatures of neutron stars on the initial cooling curve. These
pulsars [Vela (PSR 0833-45), Geminga (PSR 0630+18) and PSR's 0656+14,  1055-52]
span an age bracket of $10^4$ to $5\times 10^5$ yr.   Their X-ray luminosities 
are currently being compared with calculations for both the standard cooling and
the rapid cooling scenarios.    The surface temperature of a pulsar 
may not be entirely determined by the cooling of the neutron star, since
irradiation of its surface by  magnetopsheric X-rays~\cite{HR} could
also play a role. Thus, all
temperature estimates or ``measurements'' must be considered as upper limits.

\section{Neutron Star Structure} 

In hydrostatic equilibrium, the structure of a spherically symmetric 
neutron star is determined by the Tolman-Oppenheimer-Volkov (TOV)
equations~\cite{TolOv}:
\def\two{ \left[ 1 + {\dis \frac {P(r)}{\epsilon(r)} } \right] } 
\def\tri{ \left[ 1 + {\dis \frac {4\pi r^3P(r)}{M(r)c^2} } \right] } 
\def\four{ \left[ 1 - {\dis \frac {2GM(r)}{c^2r} } \right] }
\begin{eqnarray}
{\dis \frac {dM(r)}{dr} } &=& 4\pi r^2 \epsilon(r) 
\label{tov1} \\
{\dis \frac {dP(r)}{dr} } 
&=& - { \dis \frac {GM(r)\epsilon(r)}{c^2r^2} } 
~~ {\dis \frac {\two \tri}{\four} }\;.
\label{tov2} 
\end{eqnarray}
Above, $G$ is the gravitational constant, 
$P$ is the pressure, $\epsilon$ is the energy density inclusive of the rest
mass density, and $M(r)$ is the enclosed gravitational mass. The quantity 
$R_s=2GM/c^2$ is known as the Schwarzschild radius. 
The gravitational and baryon masses of the star are defined by 
\be
M_Gc^2 &=& \int_0^R dr\, 4\pi r^2~\epsilon(r) \\
M_Ac^2 &=& m_A\int_0^R dr\, 4\pi r^2~ 
{\dis \frac {n(r)}{\left[1 - {\dis \frac {2GM(r)}{c^2r}} \right]^{1/2} }}\;,
\ee
where $m_A$ is the baryonic mass and $n(r)$ is the baryon number density.  
The binding energy of the star is then $B.E. = (M_A - M_G)c^2$. 

By specifying the EOS of enclosed matter, $P = P(\epsilon)$, the
structure of the star is determined by choosing a central pressure
$P_c=P(\epsilon_c)$ at $r=0$ and integrating the coupled differential 
equations (\ref{tov1}) and (\ref{tov2}) out to the star surface at $r=R$ 
determined by the condition $P(r=R) =0$.  
The significance of general relativity may be gauged by the magnitude of 
$R_s/R$.  When $R_s/R  \ll 1$,  the structure is essentially
determined by Newtonian gravity.   The surface approaches the event horizon as
$R_s/R \rightarrow 1$; larger values result in black hole configurations.

\subsection{ Macroscopic properties }

A comparison of the results obtained by solving the TOV equations for   two
different EOS's which differ in their high density behavior offers
useful insights concerning the physical attributes of a star.  Figure 4 
shows some of these features.  The top left panel contains $P$ {\it vs.}
$\epsilon$ for two representative EOS's. The dash--dotted curve shows the causal
EOS $P = \epsilon$.  The solid curve is referred to as the soft EOS, since
the pressure varies less  rapidly with the energy  density than the 
dashed curve, 
which is termed the stiff EOS.   The EOS directly influences the physical
attributes of a star.  For example, the limiting or maximum mass for the stiff
EOS is larger than that for the soft EOS.  This is seen from  the mass versus
radius plots (top right panel) and also from the mass versus central baryon
density (center left panel). Also, the stiffer the EOS,
 the lower is the central
density of the maximum mass star.    Further, the radius (top right panel) and
the binding  energy (center right panel) of the maximum mass star are larger
for the stiff EOS than for the soft EOS.  Finally, the bottom panels contrast
the  moments of inertia (bottom left)  and the surface red shift (bottom right)
for the soft and stiff EOS's.   Stellar configurations with central densities
$n_c > n_c(M_{max})$   are unstable towards small perturbations, since the
gravitational attraction in such stars overwhelms the repulsive forces in
matter. Such stars are thought to form black holes.  Stars with central
densities $n_c < n_c(M_{max})$ represent stable configurations and are
candidates for stable neutron stars.  

\subsection{ Constraints on the equation of state } 

Stringent constraints  may be placed on the EOS if measurements of stars'
masses, radii, pulsar frequencies, moments of inertia, {\it etc.} are 
available. To
see how this works, it is useful to consider the dependence of some of these
properties on the mass and radius in a so-called $M-R$ diagram.   We turn now
to describe the basic elements of such an exercise. 
\begin{enumerate}
\item The denominator in Eq.~(\ref{tov2}) requires that the radius $R > R_s =
2GM/c^2$.  This yields $M/M_\odot \leq R/R_s(\odot)$, where 
$R_s(\odot)=2GM_\odot/c^2 \cong 2.95$ km. 
\item By requiring the pressure in the center of the star to be finite, 
$P_c < \infty$, Weinberg~\cite{Wein} has shown that $R >
(9/8)R_s$.  This translates to  $M/M_\odot \leq (8/9)R/R_s(\odot)$. 
\item The condition that the adiabtic sound speed $c_s=
(dP/d\epsilon)^{1/2}\leq c$, where $c$ is the speed of light, leads to the
condition that $R > 1.39R_s$, giving $M/M_\odot \leq R/(1.39\times 
R_s(\odot))$.   
\item If, instead of employing the causal EOS $P=\epsilon$ at all
densities, one requires it to hold above a fiducial density $n_t\cong 2n_0$
(below which the EOS is presumed to be known), the limit $R > 1.52R_s$ is
obtained~\cite{LPMY}, yielding $M/M_\odot \leq R/(1.52\times R_s(\odot))$.  
\end{enumerate}
Notice that the above restrictions follow from general principles.  In
practice, however, these conditions allow a large class of EOS's 
to be consistent with observations.  One is thus forced to utilize additional 
constraints that are based on observations.  This procedure, however, is
necessarily time dependent insofar as future observations may alter the 
precise limits!  In the hope that continued measurements will indeed prescribe
such limits, we will note the constraints employed currently. 

\begin{enumerate}
\item The limiting or maximum mass $M_{max}$ should exceed the largest of the
observed neutron star masses.  Currently, this condition is taken to imply
$M_{max} \geq 1.44M_\odot$, the most accurately measured neutron star
mass~\cite{mass,bww}.
\item Nearly all (up to $99\%$) of the binding energy of 
a neutron star is released in the form of neutrinos during the birth of a
neutron star, after the gravitational core collapse of a massive ($\geq
8M_\odot$)  star results in a type II supernova.  
Estimates~\cite{birth} of
the energy released in neutrinos from the SN 1987A explosion lie in the 
range $(2-4)\times 10^{53}$ ergs.  This places a restriction on the EOS that
the  ${\rm B.E} \geq (2-4)\times 10^{53}$ ergs.     
\item The Keplerian
frequency of the star (this is the rotational frequency $\Omega_K$ beyond which
the star will begin to shed mass at the equator)  should exceed the spin period
of the fastest spinning pulsar, namely that of PSR 1957+20.  This translates to
$P_K \geq 1.56~{\rm ms}$. (In reality, a star may spin at a frequency
lower than $\Omega_K$ due to viscous effects.  Choosing the larger Keplerian
frequency thus gives an upper limit.)   General relativistic calculations of
rapidly rotating stars give~\cite{LPMY} 
\be
\Omega_K \cong  7.7\times 10^3
(M_{max}/M_\odot)^{1/2}(R_{max}/10~{\rm km})^{-3/2}~{\rm s}^{-1} \,,
\ee
where $M_{max}$ and $R_{max}$ refer
to the mass and radius of the non-rotating spherical configuration.  It is
worthwhile to note that the discovery of a sub-millisecond pulsar
(say of 0.5 ms), as was purported~\cite{Kris} to be the case in the wake of
the SN1987A explosion and later retracted~\cite{Pen},
would place rather severe limits on the EOS.             
\item  Another limit~\cite{Haen,PRL} employs the
maximum moment of inertia, expressed in terms of $M_{max}$ and $R_{max}$ of the
non-rotating configurations: 
\be
I_{max} = 0.6\times 10^{45} 
\frac {(M_{max}/M_\odot) (R_{max}/10~{\rm km})^2}
{1-0.295(M_{max}/M_\odot)/(R_{max}/10~{\rm km}) }
~~{\rm g~cm^2}.
\ee
Precise measurements of $I$ would severely limit the allowed region
in the $M-R$ plane.
\end{enumerate}
In Fig. 5, the implications of the various restrictions mentioned
above are  considered.  Besides the theoretical constraints, observational
constraints imposed by mass, moments of inertia, and pulsar periods are 
illustrated.  Also shown are the mass-radius relationships of two
representative EOS's.  To date, data are consistent with a wide
variety of EOS's, which highlights the need for continuing observations.

\section{The Fate of a Newborn Neutron Star } 
 
After a supernova explosion, the gravitational mass of   
the remnant is less than 1 M$_\odot$. It is 
lepton rich and has an entropy per baryon 
of $S\simeq1$ (in units of Boltzmann's constant $k_B$).   
The leptons include both electrons and neutrinos, the latter being trapped in
the star because their mean free paths in the dense matter are of order 1 cm, 
whereas the stellar radius is about 15 km.  Accretion onto the neutron star
increases its  mass to the 1.3--1.5 M$_\odot$ range, and should mostly cease
after a second. It then takes about 10--15 s \cite{birth} for the trapped
neutrinos to diffuse out, and in the diffusion process they leave behind most
of their energy, heating the protoneutron star to fairly  uniform entropy
values of about $S=2$.   Cooling continues as thermally-produced neutrinos
diffuse out and are emitted.  After about 50 s, the star becomes completely 
transparent to neutrinos, and the neutrino luminosity drops precipitously
\cite{bur}.   
                       
Denoting the maximum mass of a cold, catalyzed neutron star by $M_{max}$ 
and the maximum mass of the protoneutron star with abundant trapped leptons
by $M_{max}^L$, there are two possible ways that a black
hole could form after a supernova explosion.  First, accretion of sufficient
material could increase the remnant's mass to a  value greater than either
$M_{max}$ or $M_{max}^L$ and produce a black hole, which then appears on the
accretion time scale \cite{bbw}.  
Second, if exotic matter plays a role  
and if accretion is insignificant after a
few seconds, then for $M_{max}^L>M>M_{max}$, where $M$ is the final 
remnant mass, a black hole will form as the neutrinos diffuse out 
\cite{prep,tpl,keiljan,pcl,subside} on the deleptonization 
time scale of 10--15 s.  
         
The existence of metastable neutron stars has some interesting implications. 
First, it could explain why no neutron star is
readily apparent in the remnant of SN1987A despite our knowledge that one
existed until at least 12 s after the supernova's explosion (see Sec. 6).  
Second, it would suggest that a significant population of relatively low mass 
black holes exists \cite{bb94}, one of which could be the compact object in the 
X-ray binary 4U1700-37 \cite{bww}.
 
How is the stellar structure, particularly the  maximum mass,
influenced by the trapped neutrinos? (The finite entropy 
plays a lesser role \cite{prep}.) 
In order to investigate this  question, one needs the 
EOS up to $\sim10$ times the baryon density encountered in the center of a
nucleus.  The EOS at such high densities is not known with any certainty. 
Nevertheless, recent work \cite{glenhyp,kapnel,gleqk} has emphasized the
possibility that  hyperons, a condensate of $K^-$ mesons, or $u$, $d$, and $s$
quarks,  may be present in addition to  nucleons and leptons.   These
additional components can appear separately or in combination with one
another.  
Compared to a star containing just plain-vanilla nucleons and
leptons, the presence of these additional  components qualitatively changes 
the way in which the  structure of the star depends upon neutrino trapping
\cite{prep}.  
                
\subsection{Equilibrium conditions}
                                                   
The composition of the star is constrained by three important physical
principles: baryon conservation, charge neutrality and beta equilibrium.  The
third exists  because the time scales of weak interactions, including those of
strangeness-violating processes,  are short compared to the dynamical time
scales of evolution (see Sec. 5).  For example, the process $p+e^-
\leftrightarrow n+\nu_e$ in equilibrium establishes the relation
\begin{eqnarray} \mu \equiv \mu_n-\mu_p = \mu_e - \mu_{\nu_e} \,, \label{beta}
\end{eqnarray} allowing the proton chemical potential to be expressed in terms
of three independent chemical potentials: $\mu_n,\mu_e$, and $\mu_{\nu_e}$. 

At densities where $\mu$ exceeds the muon mass, muons can be formed by $e^-
\leftrightarrow \mu^- + \overline\nu_\mu + \nu_e$, hence the muon chemical
potential is  \begin{eqnarray} \mu_\mu=\mu_e-\mu_{\nu_e}+\mu_{\nu_\mu}\,, 
\end{eqnarray} requiring the specification of an additional chemical potential
$\mu_{\nu_\mu}$. However, unless $\mu > m_\mu c^2$, the net number of $\mu$'s
and $\nu_\mu$'s per baryon (designated by $Y_\mu$ and $Y_{\nu_\mu}$,
respectively) is zero, because no muon-flavor leptons are present at the onset
of trapping, so $Y_{\nu_\mu}=-Y_\mu$ determines $\mu_{\nu_\mu}$.  Following
deleptonization, $Y_{\nu_\mu}=0$, and $Y_\mu$ is determined by $\mu_\mu=\mu_e$
for $\mu_e > m_\mu c^2$, or by $Y_\mu=0$ otherwise.           
                            
Negatively charged kaons can be formed in the process
$n+e^- \leftrightarrow n + K^-+\nu_e$ when $\mu_{K^-} = \mu$
becomes equal to the energy of the
lowest eigenstate of a $K^-$ in matter.
In addition, weak reactions for the $\Lambda, \Sigma$, and $\Xi$
hyperons are all of the form $B_1+\ell \leftrightarrow B_2+\nu_\ell$, where
$B_1$ and $B_2$ are baryons, $\ell$ is a lepton, and $\nu_\ell$ is a neutrino of
the corresponding flavor. The chemical potential for a baryon $B$ with baryon
number $b_B$ and electric charge $q_B$ is then given by the general relation
\begin{eqnarray}
\mu_B=b_B\mu_n - q_B\mu \,, \label{echem}
\end{eqnarray}
which leads to 
\begin{equation}
\mu_{\Lambda}= \mu_{\Sigma^0} = \mu_{\Xi^0} = \mu_n \quad;\quad
\mu_{\Sigma^-}= \mu_{\Xi^-} = \mu_n+\mu\quad;\quad
\mu_p = \mu_{\Sigma^+} = \mu_n - \mu\,. \label{murel}
\end{equation}
The same considerations apply to quarks, for which Eq.~(\ref{echem}) gives
\begin{equation}
\mu_d = \mu_s = (\mu_n+\mu)/3 \quad;\quad
\mu_u = (\mu_n-2\mu)/3 \,.
\end{equation}

Therefore, if there are no trapped neutrinos present, so that     
$\mu_{\nu_\ell}=0$, there are two independent chemical potentials ($\mu_n,
\mu_e$) representing conservation of baryon number and charge.  If trapped
neutrinos are present ($\mu_{\nu_\ell}\neq 0$), further constraints, due to
conservation of the various lepton numbers over the dynamical time scale of
evolution, must be specified.  At the onset of trapping, during the initial
inner core collapse, $Y_\ell=Y_e+Y_{\nu_e}\approx 0.4$ and $Y_e/Y_{\nu_e} \sim
5-7$, depending upon the density \cite{pcollapse}.  These numbers are not
significantly affected by variations in the EOS.   Following deleptonization,
$Y_{\nu_e}=0$, and $Y_e$ can vary widely, both with the density and EOS.  
                                                                               
As long as both weak and strong interactions are in equilibrium,  the above
general relationships determine the constituents of the star during its
evolution.  Since electromagnetic interactions give negligible
contributions, it is sufficient to consider the non-interacting (Fermi gas)
forms for the partition  functions of the leptons.  Hadrons, on the other hand,
receive significant contributions at high density from the less well known 
strong interactions.

\subsection{Neutrino-poor versus neutrino-rich stars } 

A detailed discussion of the composition and structure of protoneutron stars
may be found in Refs. \cite{prep,comments}.   The main findings were that
the structure depends more sensitively on the compostion of the star than its
entropy  and that  the trapped neutrinos play an important role in determining
the composition.  Since the structure is chiefly determined by the pressure of
the strongly interacting constituents and the nature of the strong interactions
is poorly understood at high density, several models of dense matter,
including  matter with strangeness-rich hyperons, a kaon condensate and quark
matter were studied there.  For the purpose of illustration, 
we show here only the cases in which 
hyperons and quarks appear at high densities. 
The qualitative trends when other forms of strangeness, including kaon
condensates, appear are very similar.

Figure 6 shows the various concentrations, 
$Y_i$ (the number of particles of species $i$ per baryon), as a 
function of density when the only
hadrons allowed are nucleons.  The arrows indicate the central density of
the maximum mass stars. The left hand panel refers to the case in which
the neutrinos have left the star. At high density the proton concentration is
about 30\%, charge neutrality ensuring an equal number of negatively charged
leptons. This relatively large value is the result of the symmetry energy
increasing nearly linearly with density in this model.  Many non-relativistic 
potential models \cite{wff} predict a maximum proton concentration of only
10\%. However, the generic results discussed here are not sensitive to the
behavior of the nuclear symmetry energy.  The effects of neutrino trapping are
displayed in the right hand panel. The fact that $\mu_{\nu_e}\neq0$ in Eq.
(\ref{beta}) results in larger values for $\mu_e$ and $Y_e$. Because of charge
neutrality, $Y_p$ is also larger, and it approaches 40\% at high density.  It
is clear that the maximum mass configuration has a much lower density when
neutrinos are trapped than when they are not.  As is evident from 
the bottom row of Table 1,   
neutrino-trapping reduces the maximum mass $M_{max}^L$ from the value found in
neutrino-free matter $M_{max}$; although neutrino-trapped nucleons-only matter
contains more leptons and more leptonic pressure, it also contains more protons
and, therefore, less baryonic pressure.   Thermal effects increase the pressure
and therefore the maximum mass, but only slightly.  Even for $S=2$, the central
temperature is only $\sim50$ MeV, which is  much less than the  nucleon Fermi
energies. Thus, because $M_{max}^L\lord M_{max}$, a black
hole could only form promptly after bounce from nucleons-only stars, 
in the absence of significant accretion at late times.  
 
\begin{quote}                                  
{{\bf Table  1.}  Maximum masses of stars with  baryonic matter that undergoes a phase
transition to quark matter without ($Y_\nu=0$) and with ($Y_{Le}=0.4$) trapped
neutrinos.  Results are for a mean field model of baryons and a bag model of
quarks. $B$ denotes the bag pressure in the quark EOS. }
\end{quote}
\begin{center}
\begin{tabular}{c|c|c|c|c}
\multicolumn{5}{c}{$M_{\rm max}/M_{\odot}$} \\ \hline \hline
$ \begin{array}{c} \, ~ \, \\ \, ~ \, \end{array}$ 
& \multicolumn{2}{c|}{Without hyperons} 
& \multicolumn{2}{c}{With hyperons} \\ \hline
$ \begin{array}{c} B \\ ({\rm MeV~fm}^{-3}) \end{array} $ 
& $Y_{\nu}=0$ & $Y_{Le}=0.4 $ & $Y_{\nu}=0$ & $Y_{Le}=0.4$ \\ \hline
136.6    & 1.440 & 1.610 & 1.434 & 1.595 \\
150      & 1.444 & 1.616 & 1.436 & 1.597 \\
200      & 1.493 & 1.632 & 1.471 & 1.597 \\
250      & 1.562 & 1.640 & 1.506 & 1.597 \\ \hline
${\rm No~~quarks}$ & 1.711 & 1.645 & 1.516 & 1.597 \\ \hline \hline
\end{tabular}
\end{center}

For comparison, Fig. 7 shows the compositions  of neutrino-free matter 
(left panel) and neutrino-trapped matter (right panel) in the event that 
strange particles are allowed to appear.  In neutrino-free matter, one expects 
that the $\Lambda$, with a mass of 1116 MeV, and the $\Sigma^-$, with a
mass of 1193 MeV, both first appear at roughly the same density, because the
somewhat higher mass of the $\Sigma^-$ is compensated by the presence of
$\mu_e$  in the equilibrium condition of the $\Sigma^-$.  More massive and more
positively charged particles than these appear at higher densities.  Notice
that with the appearance of the negatively charged $\Sigma^-$ hyperon, which
competes with the leptons in maintaining charge neutrality, the lepton
concentrations begin to fall.  When quarks appear, at 
around $4n_0$ (for $B=200~{\rm MeV~fm^{-3}}$),
 the neutral and negative particle
abundances begin to fall, since quarks furnish both negative charge and baryon
number.  

Trapped neutrinos again increase the proton and electron abundances, and this
strongly influences the threshold for the appearance of hyperons.  The
$\Lambda$ and the $\Sigma$'s now appear at densities higher than those found in
the absence of neutrinos.  In addition, the transition to a mixed phase with
quarks is delayed to about  $10n_0$, which is beyond the central density
of the maximum mass star (see arrow in Fig. 7). 
This makes the overall EOS stiffer, so that
when matter contains strangeness the behavior is opposite to that
of the nucleons-only case.  Specifically, the
maximum mass is {\it larger} in the
neutrino-trapped case.  This behavior is summarized in Table~1 for
different assumptions about the composition and was first noted by Thorsson,
Prakash and  Lattimer~\cite{tpl}  in the study of stars with kaon condensates
and subsequently by Keil and Janka~\cite{keiljan} of stars with hyperons.

Evolutionary calculations \cite{birth,keiljan} without accretion show that it
takes on the order of 10--15 s for the trapped neutrino fraction to     vanish
for a nucleons-only EOS 
(see also Sec. 5). 
To see qualitatively what might transpire during the early evolution, we show 
in Fig. 8 the dependence of the maximum stellar mass upon the trapped neutrino
fraction $Y_{\nu_e}$, which decreases during the evolution. When the only
hadrons are nucleons (np), the maximum mass increases with 
decreasing $Y_{\nu_e}$, whereas 
when hyperons (npH) or kaons (npK) are also present, it decreases. Further, the
rate of decrease accelerates for rather small values of $Y_{\nu_e}$.  Coupled
with this is the fact that the central density of stars will tend to increase
during deleptonization.  The implication is clear. {\it If} ~hyperons, kaons,
or other negatively-charged hadronic species are present, an initially
stable star can change into a black hole after most of the trapped neutrinos
have left, and this takes $10-15$ s.  This happens only if the remnant mass $M$
satisfies $M_{max}^L>M>M_{max}$.

It must be emphasized that the maximum mass of the cold catalyzed star still
remains uncertain due to the uncertainty in strong interactions at high
density. At present, all nuclear models can only be effectively constrained at
nuclear density and by the condition of causality at high density. The
resulting uncertainty is evident from the range of possible maximum masses
predicted by the different models.  Notwithstanding this uncertainty, our
findings concerning the effects of neutrino trapping offer intriguing
possibilities for distinguishing between the different physical states of
matter.  These possibilities include both black hole formation in supernovae
and the signature of neutrinos to be expected from supernovae. 

\section{Evolution Time Scales } 

The evolution of the star will be governed by the time scale for neutrino
diffusion, whether the star is in the deleptonizing or cooling phase. These
time scales depend on the EOS, which determines the stellar size, and the
neutrino opacities.  The composition and temperature in the central regions of
the star at the beginning of deleptonization are characterized by a high lepton
fraction ($Y_{Le}=0.4$) and low entropy per baryon ($S \sim 1$), while the
cooling phase is characterized by $Y_{\nu_e} \sim 0$ and $S \sim 2$.  A full
treatment of neutrino transport during the evolution requires opacities for a
wide range of composition and matter degeneracy. The dynamical changes in the
lepton fraction and the temperature modify the composition of matter and the
typical neutrino energies in the inner core.  Roughly, the neutrino diffusion
time scale is proportional to $R^2(c\lambda)^{-1}$, where $R$ is the stellar
radius and $\lambda$ is the appropriate effective neutrino mean free path.   In
Ref. \cite{prep}, the time scales in the two major evolutionary phases were
estimated for nucleons-only matter,  using semi-analytical considerations.  In
the deleptonization phase, the diffusion time is dominated by charged current
interactions and was found to be      
\begin{eqnarray} 
\tau_d \simeq \frac {9R^2}{19c\lambda_{abs}^o} \simeq \frac
{44.3}{f}  \left ( \frac {R}{10 ~{\rm km}} \right)^2 \, \rm s \,, \label{taud}
\end{eqnarray} 
where $\lambda_{abs}^o$ is the fiducial absorption mean free path at the
expected initial $\nu_e$ energy, $E_{\nu_e} \simeq 260$ MeV.   The factor
$f=3-10$ accounts for degeneracy and interaction effects and is not yet
well-determined.  This leads to a deleptonization time of 5-15 s, in agreement
with more detailed numerical simulations.  The result in Eq. (\ref{taud}) shows
clearly how the deleptonization time depends on the EOS through the radius of
the star, $R$, and the opacity. Similarly, the thermal cooling time, in which
neutral current interactions are also important, was estimated to be $\tau_c =
(1-2)\tau_d$.

It is of some importance to contrast these dynamical time scales with the weak
interaction time scales controlling the onset of strangeness-bearing
components.  We have seen earlier that strangeness is most likely to appear
near the end of deleptonization.  During this epoch, the central temperatures
are expected to be high ($T= 10-30$ MeV).  The time scale for the onset of
strangeness in the form of a kaon condensate was recently addressed by Muto
{\it et al.}  \cite{MTI}, who considered the time evolution of the condensate
amplitude $\theta(t)$, but only at $T=1$ MeV, which would be appropriate for a
cold catalyzed star.  They found that the onset of strangeness was of the
same order as the cooling time, which would have major implications for the
neutrino luminosity of a cooling neutron star.  The relevant weak interaction
rates are, however, extremely sensitive to temperature. Figure 9 shows the time
evolution of the condensate amplitude close to the threshold density for
temperatures in the range 1--10 MeV.  It is evident that, for temperatures
appropriate for the cooling phase, the weak interaction time scales are much
smaller than the duration of the cooling phase.  Therefore, to a very good
approximation, the appearance of a kaon condensate may be considered as
instantaneous.  The onset of strangeness in the form of hyperons and/or quarks
is likely to be similarly rapid.  

\section{Supernova SN1987A}  

\noindent On February 23 of 1987, neutrinos were observed \cite{exp87a} from
the explosion of supernova SN1987A, indicating that a neutron star, not a
black hole, was initially present.   (The appearance of a black hole  would
have caused an abrupt cessation of any neutrino signal \cite{bur}.)  The
neutrino signal was observed for a period of at least 12 s, after which
counting statistics fell below measurable limits.  From the handful of events
observed, only the average neutrino energy,  $\sim 10$ MeV, and the  total
binding energy release of $\sim (0.1-0.2){\rm M}_\odot$ could be estimated.  
                                                      
These estimates, however, do not shed much light on the composition of the
neutron star.  This is because, to lowest order, the average neutrino energy
is fixed by the neutrino mean free path in the outer regions of the
protoneutron star. Further, the binding  energy exhibits a  universal
relationship \cite{prep} for a wide class of EOS's, including those with
strangeness  bearing components, namely 
\begin{eqnarray}
B.E. = (0.065\pm 0.01)(M_B/{\rm M}_\odot)^2{\rm M}_\odot \,, 
\end{eqnarray}
where $M_B$ is the baryonic mass. This  allows us only to determine a remnant
gravitational mass of $(1.14-1.55){\rm M}_\odot$, but not the composition.   

The ever-decreasing optical luminosity (light curve) \cite{lcurve}
of the remnant of SN1987A suggests
two arguments against the continued 
presence of a neutron star.  First,  accretion onto a neutron star at 
the Eddington limit is already ruled out for the usual hydrogen-dominated 
Thomson electron scattering opacity.  (However, if the atmosphere surrounding 
the remnant contains a sufficient amount of 
iron-like elements, as Chen and Colgate \cite{cc} suggest, the
appropriate Eddington limit is much lower.) 
Second, a Crab-like pulsar cannot exist in SN1987A, since the
emitted magnetic dipole radiation would be observed in the light curve. 
Either the magnetic field or
the spin rate of the neutron star remnant would have to be much less than in 
the case of the Crab and what is inferred from other young neutron
stars.  The spin rate of a newly formed
neutron star is expected to be high; however, the time scale for the 
generation of a significant magnetic field is not well known and
could be greater than 10 years.  
                                                               
Although most of the binding energy is released during the initial accretion 
and  collapse stage in about a second after bounce,  the neutrino signal
continued for a period of at least 12 s.   The compositionally-induced changes
in the structure of the star occur on the deleptonization time scale, which we
have estimated to be of order 10--15 s \cite{prep}, not on the binding
energy release time scale.   Thus, the duration of the neutrino signal from
SN1987A was  comparable to the time required for the neutrinos initially
trapped in the star to leave.  However,  counting statistics prevented
measurement of a longer duration, and this unfortunate happenstance prevents
one from distinguishing a model in which negatively-charged matter appears and
a black hole forms from a less exotic model, in which 
a neutron star still exists.  
As we have pointed out, the maximum stable mass drops by as much as 
$0.2{\rm M}_\odot$ when the trapped neutrinos depart if negatively charged
particles are present,  which could be enough to cause collapse to a black
hole.  
                                                             
Observed neutron stars lie in a  very small range of  gravitational masses (see
Fig. 2).  Thielemann {\it et al.} \cite{thm} and Bethe and
Brown \cite{bb} have estimated the gravitational mass of the remnant of SN1987A
to  be in the range $(1.40-1.56){\rm M}_{\odot}$, using arguments based on the
observed amounts of ejected $^{56}$Ni and/or the total explosion energy.  This
range extends above the largest accurately known value for a neutron star mass, 
$1.44~{\rm M}_\odot$, so the possibility  exists that the neutron star
initially produced in SN1987A could be unstable  in the cold, deleptonized 
state. In this case, SN1987A would have become a  black hole once it had
deleptonized, and no further signal would be expected. Should this scenario be
observationally verified, it would provide strong  evidence for the appearance
of strange matter.

\section{Future Directions }

\noindent The emitted neutrinos, of all flavors, are the only direct probe of 
the mechanism of supernova explosions and the structure of newly formed 
neutron stars.  The cooling of the star can yield information on 
the stellar composition.  The two most important microphysical ingredients 
for detailed simulations of the cooling of a newborn neutron star 
are the EOS of dense matter and the neutrino opacities.  

Future efforts must address the crucial question of the strong interactions of
strange  particles in dense  matter  --  even near nuclear equilibrium density, 
our  knowledge is sketchy at present.  

Surprisingly little attention has been paid to the effects of composition and
of strong interactions of the ambient matter on neutrino opacities.  It is
essential that the opacities be consistent with the composition, which has not
been a feature of protoneutron star models to date.  A first effort in this
direction, including the presence of strangeness in the form of hyperons, has
revealed significant modifications of the neutrino opacities \cite{RP}.  
Results from this study on neutral-current neutrino cross sections are shown 
in Fig. 10.

Significant contribution to the neutrino opacity arises from scattering
involving  $\Sigma^-$ hyperons.  Although the lowest order tree level
contributions from the  neutral $\Lambda$ and $\Sigma^0$ vanish (due to their
zero weak hyper-charge), these
particles, when present, furnish baryon number, which decreases the relative
concentrations of nucleons.  This leads to significant reductions in the 
opacity.  The neutrino cross sections depend sensitively on the Fermi momenta
and effective masses of the various particles present in matter.   As long as
one or the other hyperon is present, the cross sections are significantly
modified from the case of nucleons-only matter.  
                                                                    
In neutrino-trapped matter, the appearance of negatively charged hyperons
($\Sigma^-$) is delayed to higher densities (relative to neutrino-free matter);
also, their abundances are suppressed.  However, the presence of neutral
hyperons, such as the $\Lambda$, results in neutrino abundances that grow with
density. This leads to large enhancements in the cross sections for neutrinos
(of characteristic energies close  to the local neutrino chemical potential)
compared to those in normal nucleonic matter. 
                   
Modifications due to strangeness in the cooling phase are somewhat different
than in the deleptonization phase.  In the cooling phase, in which matter is
nearly neutrino free, the response of the $\Sigma^-$ hyperons to thermal 
neutrinos is significant.  However, the presence of a large number of $\Lambda$
hyperons, to which the neutrinos do not couple, decreases the total cross
sections relative to nucleonic matter.                                     

These findings suggest several directions for further study.  These include
extension to consider correlations between the different particles and RPA
corrections.  The presence of charged   particles, such as the $\Sigma^-$,
could make available low energy collective  plasma modes through
electromagnetic correlations, in addition to the scalar,  vector, 
and iso-vector
correlations.  Calculations of neutrino opacities from  charged current
reactions (which are important during the deleptonization phase), in
strangeness-rich matter whose constituents exhibit varying degrees of
degeneracy, are required for a complete description of the evolution.  
                                
What can be expected in future detections?  In an optimistic scenario, several 
thousand neutrinos from a typical galactic supernova might be seen in  upgraded
neutrino detectors, such as SNO in Canada and Super  Kamiokande in Japan. (For
rough characteristics of present and future neutrino  detectors, see
Ref. \cite{bkg}.) Among the interesting features that could be sought are:   
\begin{enumerate}
\item Possible cessation of a neutrino signal, due to black hole formation.
\item Possible burst or light curve feature associated with the onset of
negat\-ively-charged, 
strongly interacting matter near the end of deleptonization,
whether or not a black hole is formed. 
\item Identification of the deleptonization/cooling epochs by changes in
luminosity evolution or neutrino flavor distribution.
\item Determination of a radius-mean free path correlation from the luminosity
decay time or the onset of neutrino transparency.
\item Determination of the neutron star mass from the universal  binding
energy-mass relation.
\end{enumerate}

\vskip 10pt
\noindent \large {\bf Acknowledgement}\normalsize
\vskip 10pt
\noindent
We dedicate this article to Gerry Brown in honor of his 70th birthday. 
We gratefully acknowledge collaborations with I. Bombaci, Manju Prakash, R.
Knorren, and J. R. Cooke.  This work was  supported by the U. S. Dept. of Energy
under grants DOE/DE/FG02-87ER-40388, 40328, and 40317, and  by the NASA grant
NAG52863.

\vfill\eject

\vfill\eject 

\vspace*{10cm}
\includegraphics{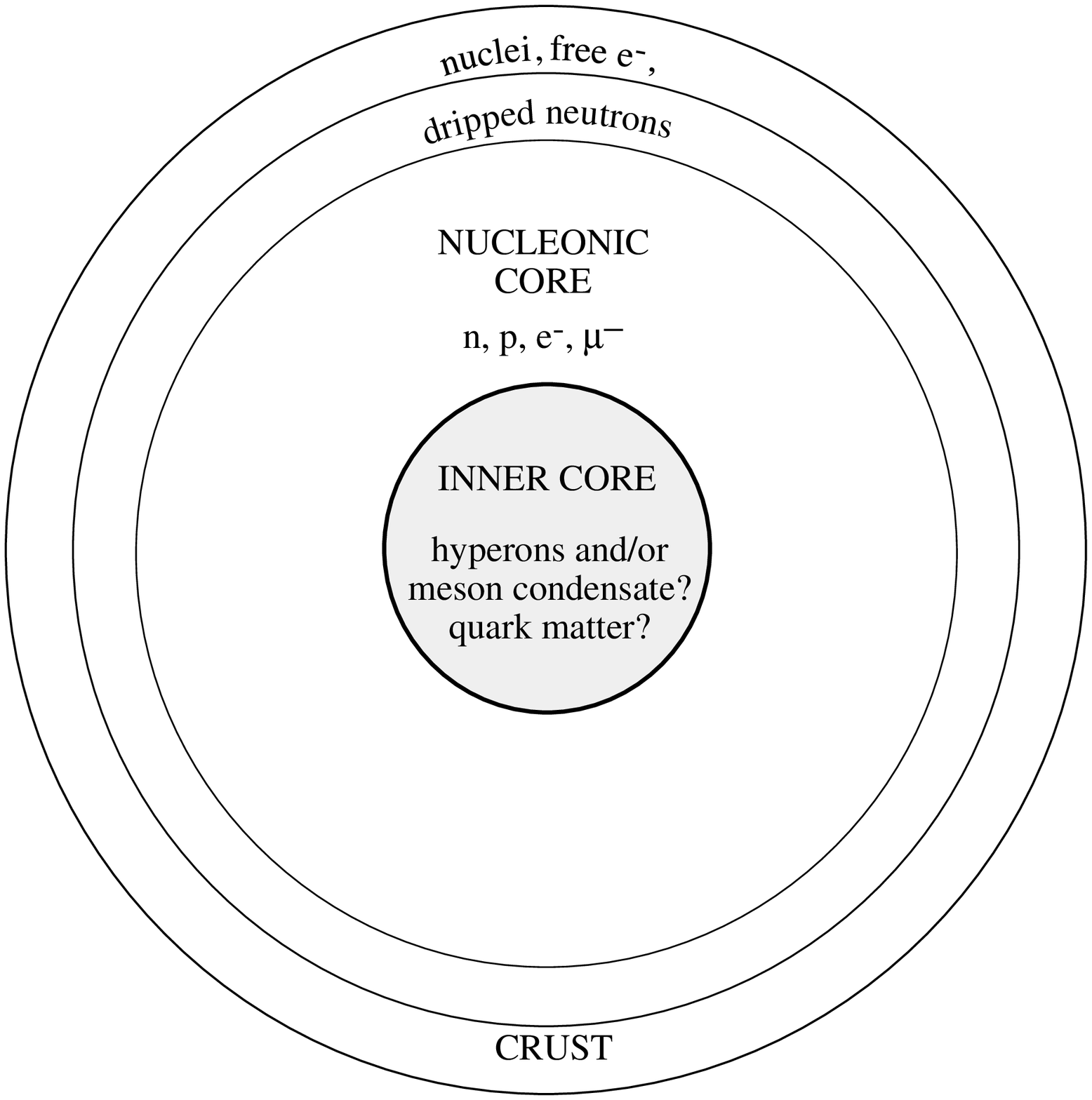}
\vskip -40pt
\begin{minipage}[t]{10cm}
\noindent \bf Fig.1.  \rm 
Schematic cross-sectional view of a neutron star.
\end{minipage}
\vskip 4truemm

\vspace*{12cm}
\includegraphics{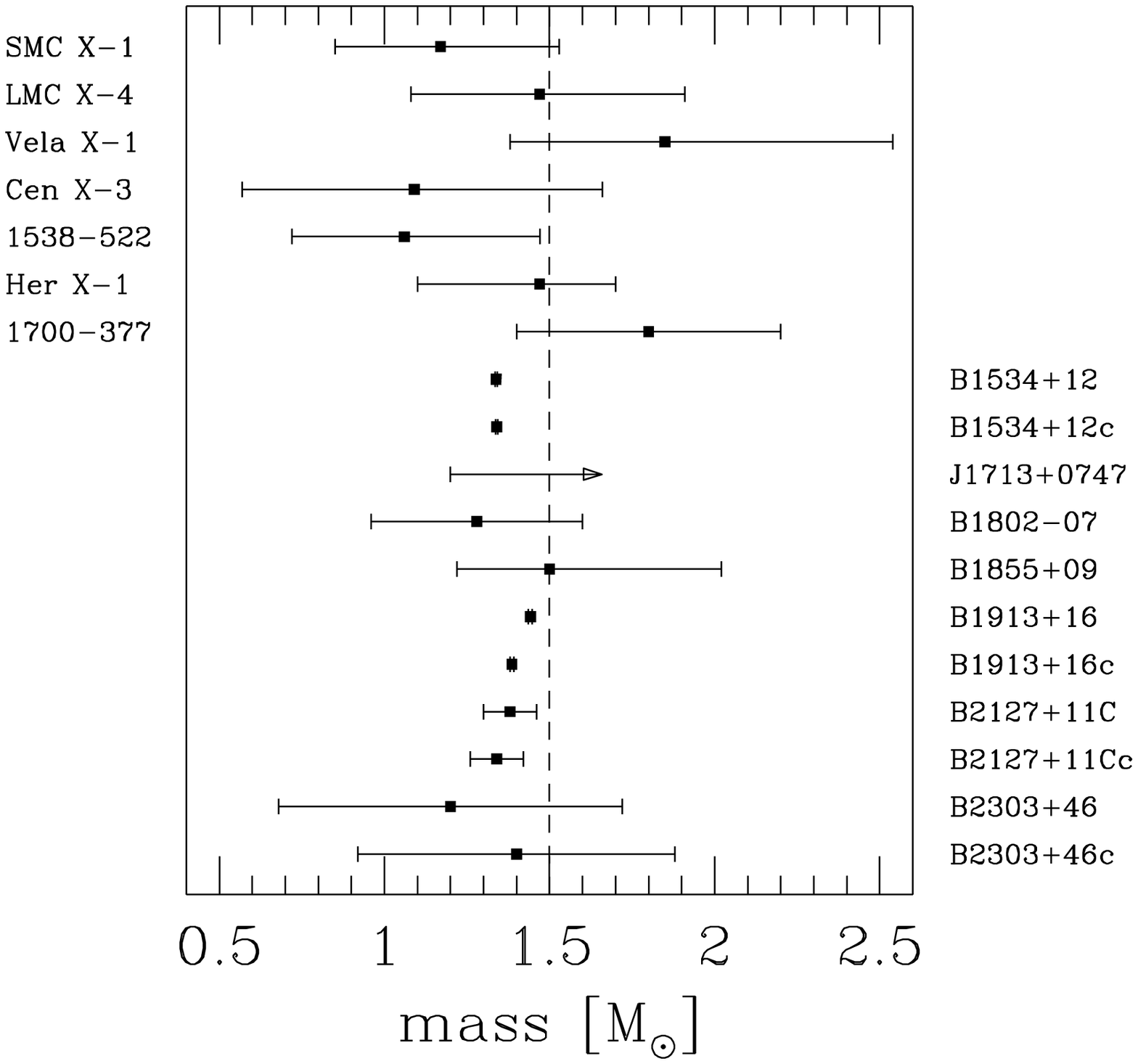}
\vskip -80pt
\begin{minipage}[t]{10.5cm}
\noindent \bf Fig.2.  \rm 
Measured neutron star masses from Refs. \cite{mass,bww}. The upper seven 
values are for X-ray pulsars, the lower eleven for radio pulsars and their 
companions. Error bars indicate 95\% confidence limits.       
\end{minipage}
\vskip 4truemm

\vspace*{13cm}
\includegraphics{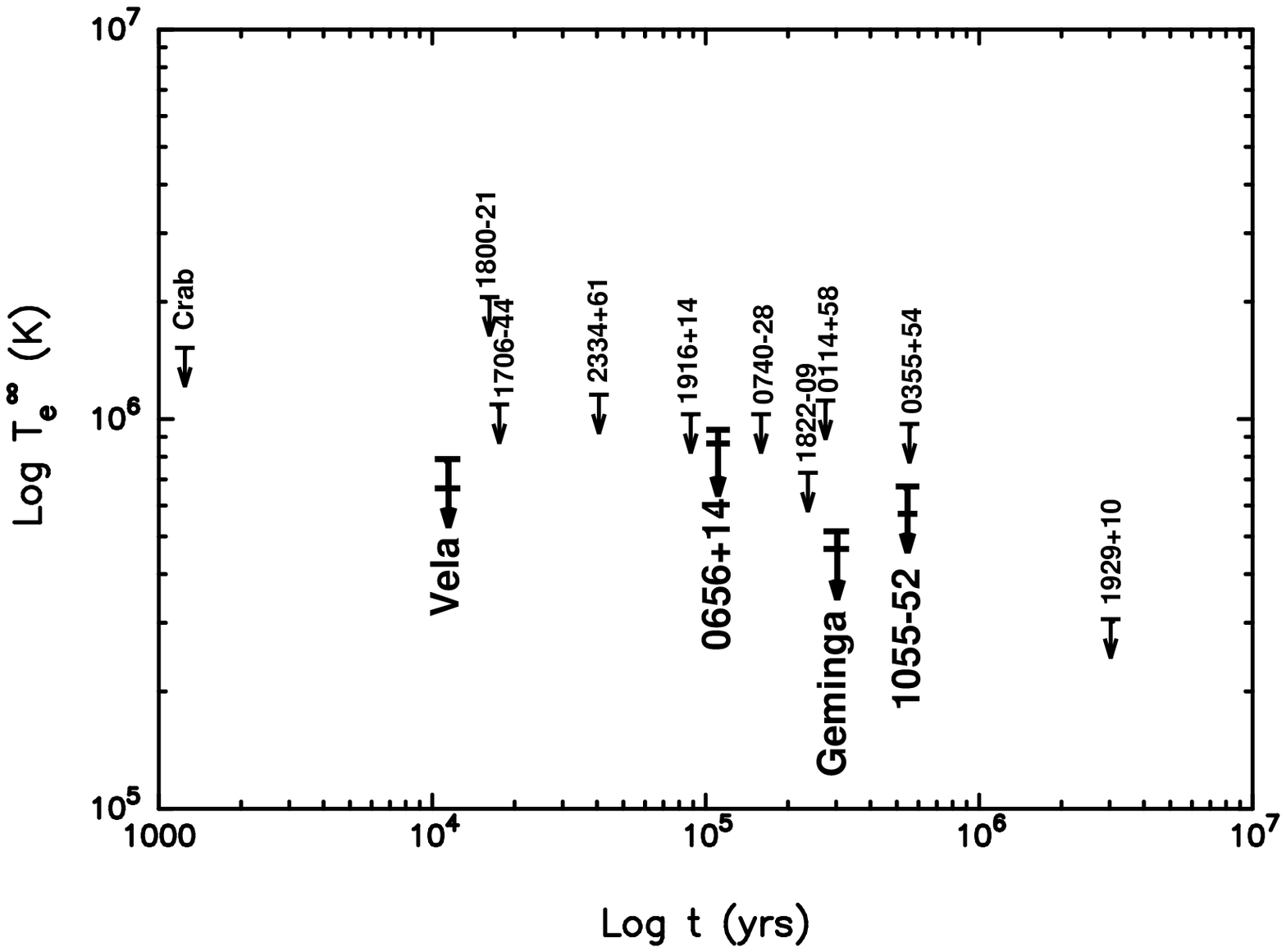}
\vskip -140pt
\begin{minipage}[t]{10.5cm}
\noindent \bf Fig.3.  \rm 
Measurement, estimates, or upper limits for the surface  temperature
of fourteen pulsars \cite{Ogel}. (Figure after Dany Page.) 
The plotted ages are the pulsar spin-down ages. 
\end{minipage}
\vskip 3truemm

\vspace*{15cm}
\includegraphics{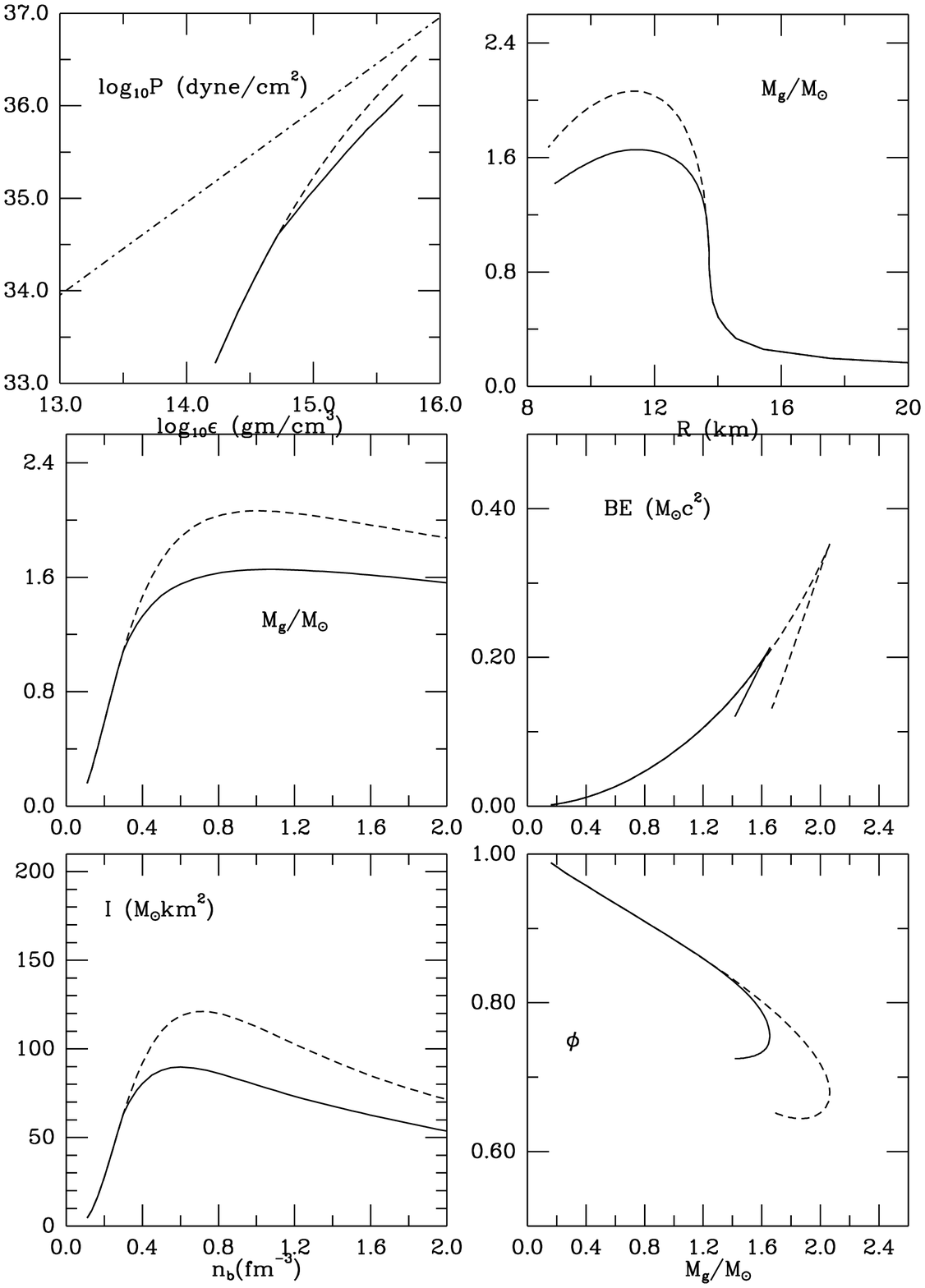}
\vskip -30pt
\begin{minipage}[t]{10cm}
\noindent \bf Fig.4.  \rm 
Stellar properties for two representative EOS's. 
\end{minipage}
\vskip 4truemm

\vspace*{10cm}
\includegraphics{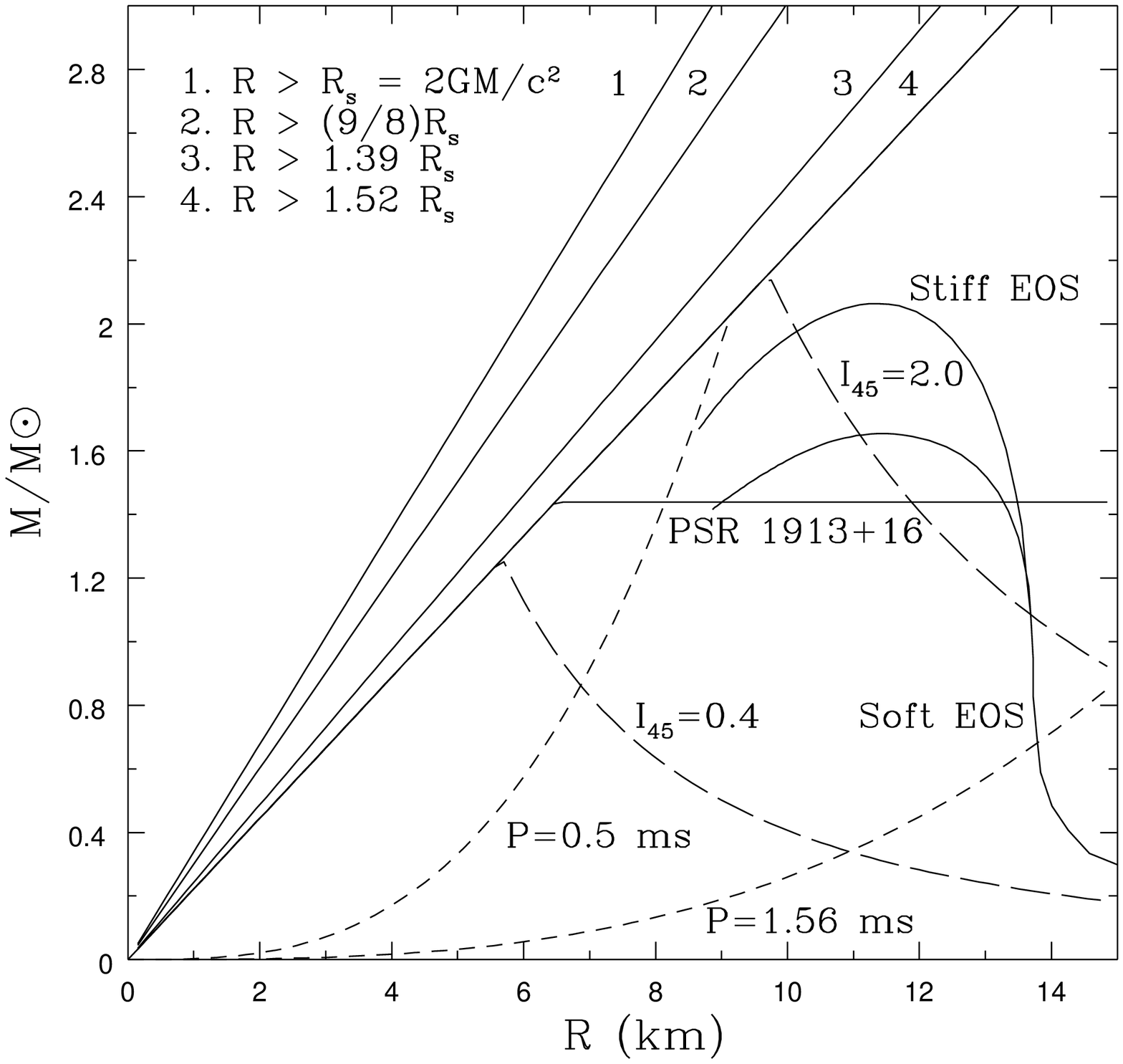}
\vskip -60pt
\begin{minipage}[t]{10cm}
\noindent \bf Fig.5.  \rm 
Valid EOS's must produce neutron stars which lie to the
right of curve labeled 4. Current observational constraints include $M > M_{PSR
1913+16},~I > 0.4\times 10^{45}~{\rm g~cm^2}$ (Crab pulsar), and $P_{min} < 1.6$
ms (PSR 1957+20). The promise of future observations is illustrated with curves
of other values of $I$ and $P$. 
\end{minipage}
\vskip 4truemm

\vspace*{14cm}
\includegraphics{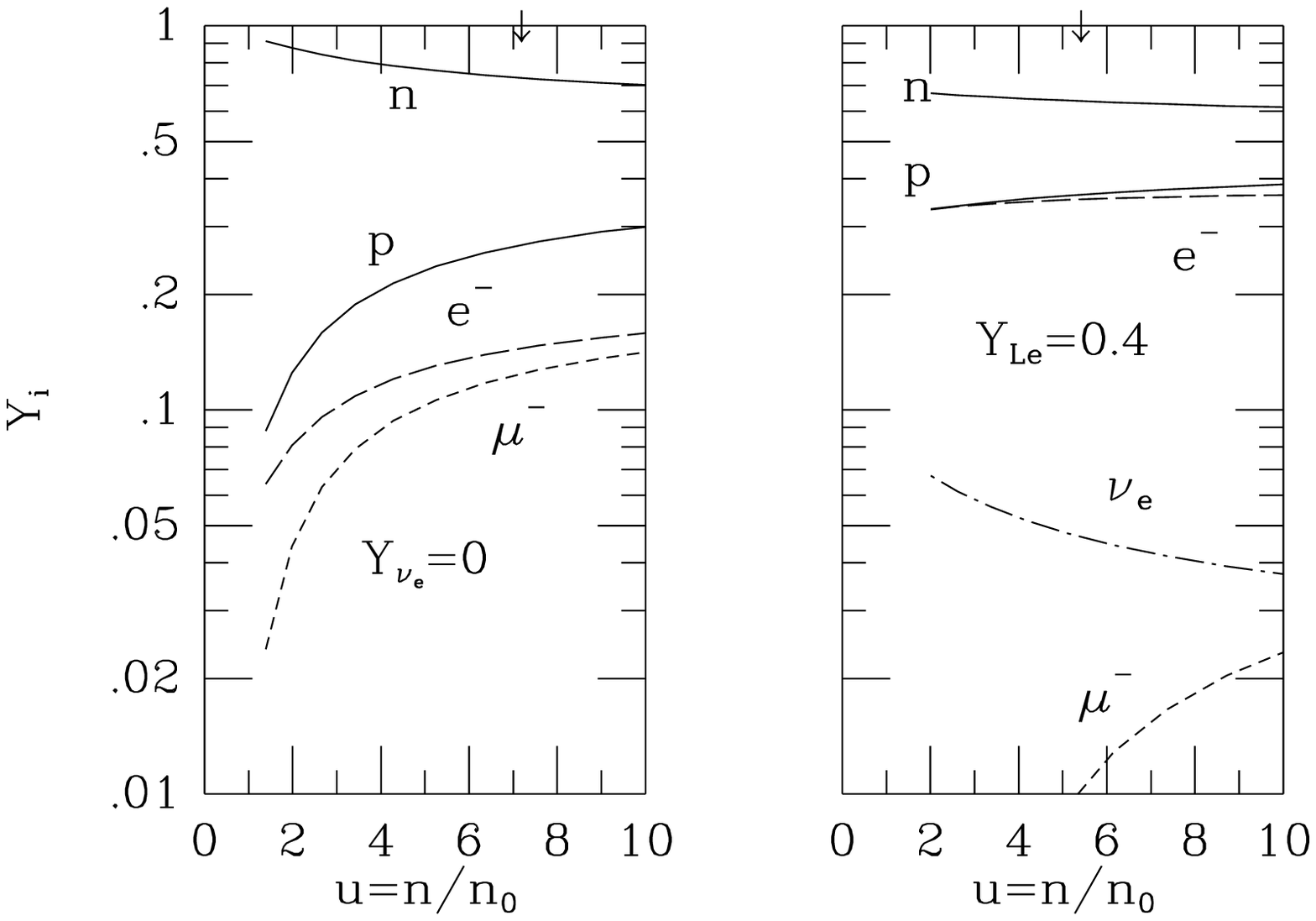}
\vskip -170pt
\begin{minipage}[t]{10cm}
\noindent \bf Fig.6.  \rm 
Individual concentrations, $Y_i$, as a function of the
baryon density ratio $u=n/n_0$, where $n_0$ is the density of 
equilibrium nuclear matter. The arrows indicate the central density of the
maximum mass stars. Left panel: neutrino free. Right 
panel: with trapped neutrinos ($Y_{\ell}=0.4$). 
\end{minipage}
\vskip 4truemm

\vspace*{12.5cm}
\includegraphics{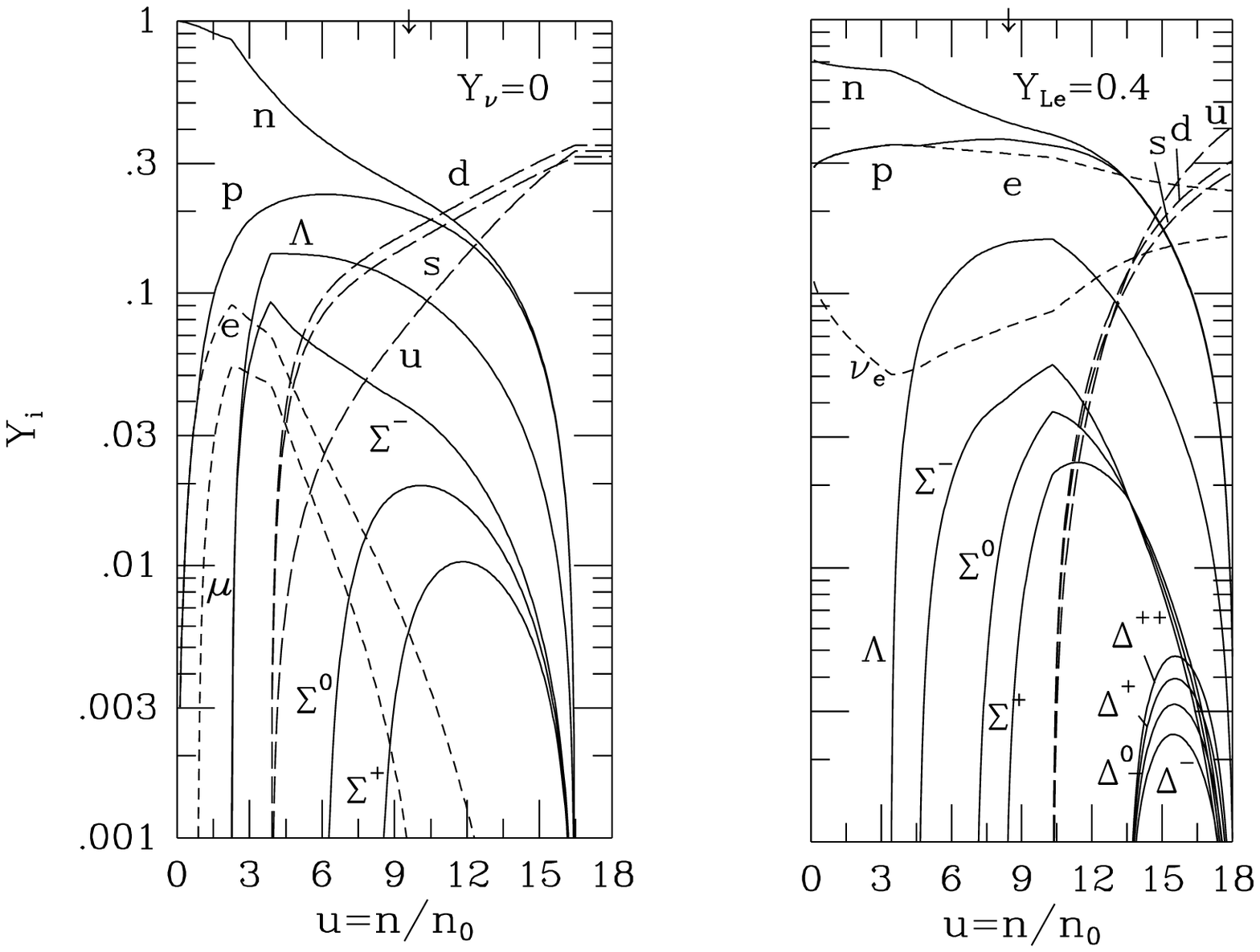}
\vskip -110pt
\begin{minipage}[t]{10cm}
\noindent \bf Fig.7.  \rm 
As for Fig. 6, but for matter which contains strangeness-bearing hyperons and 
quarks, as well as nucleons and leptons. 
\end{minipage}
\vskip 4truemm

\vspace*{9cm} \includegraphics{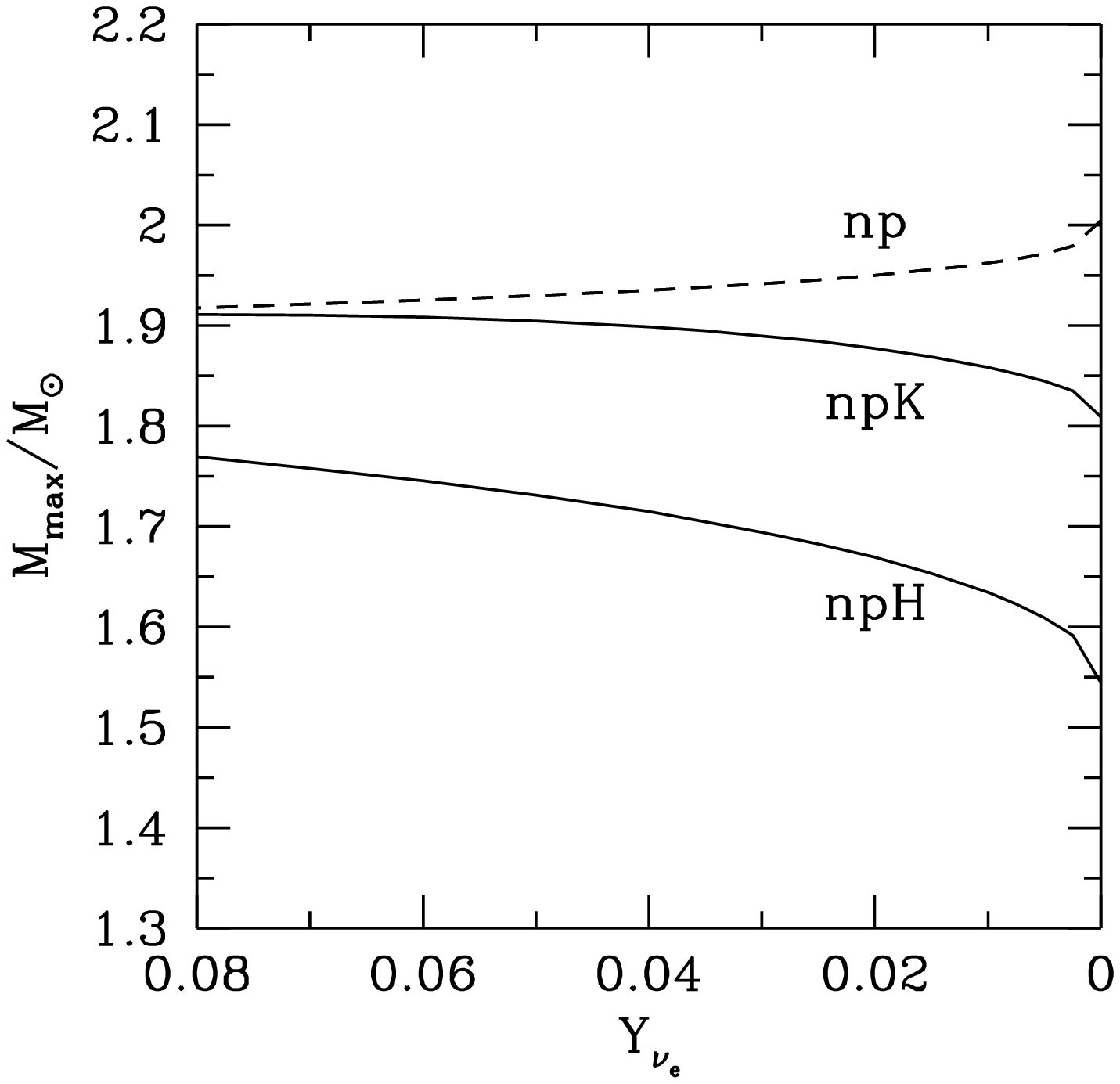}
\vskip -80pt
\begin{minipage}[t]{10cm}
\noindent \bf Fig.8.  \rm 
Maximum neutron star mass as a function of $Y_{\nu_e}$ for hadronic matter with
only nucleons (np) or with nucleons and hyperons (npH) or kaons (npK).
\end{minipage}
\vskip 4truemm

\newpage
\vspace*{10cm} \includegraphics{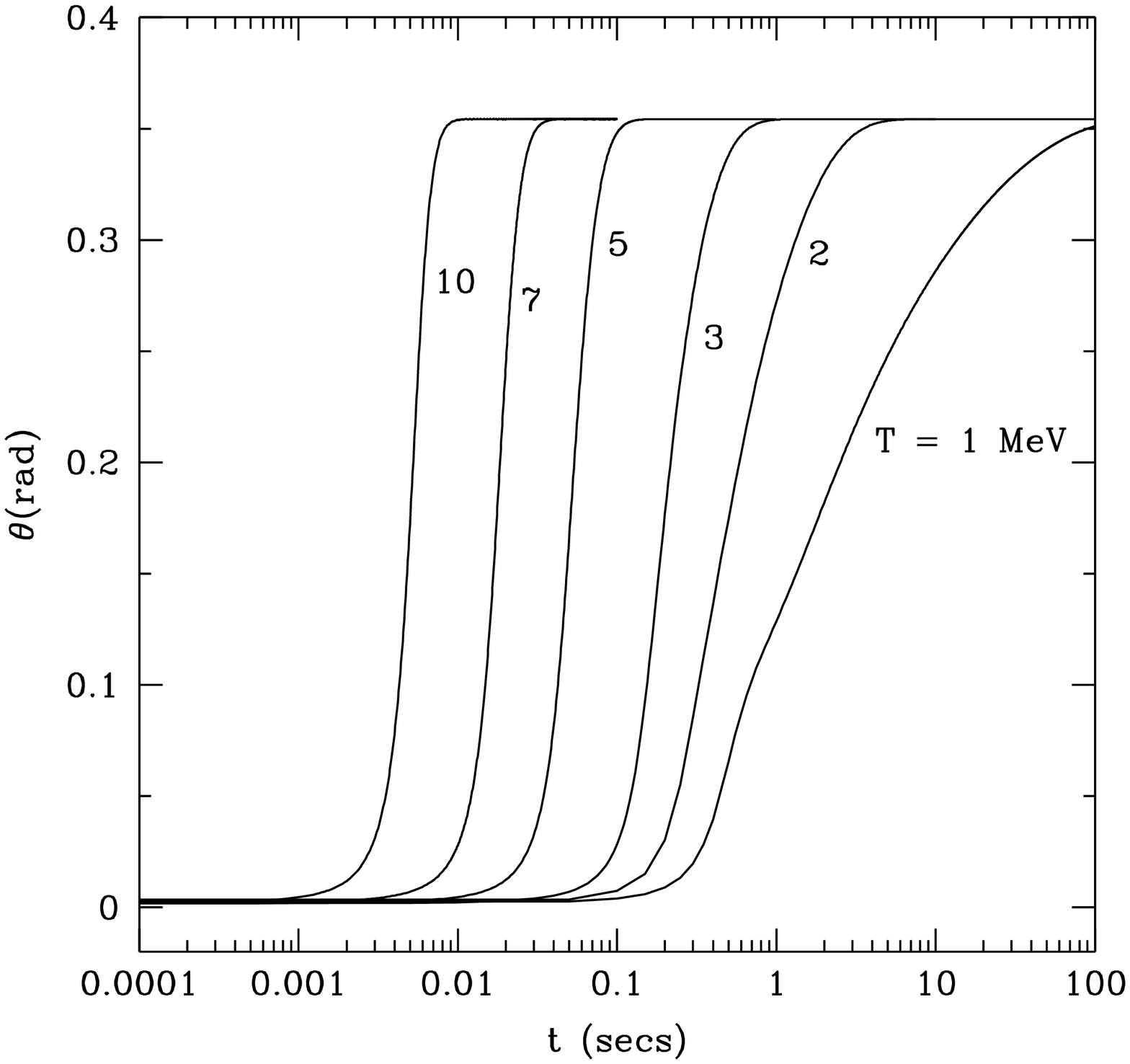} \vskip -40pt 
\begin{minipage}[t]{10cm} 
\noindent \bf Fig.9. 
\rm  Kaon condensate amplitude as a function of time.  \end{minipage} \vskip
4truemm

\vspace*{13cm}
\includegraphics{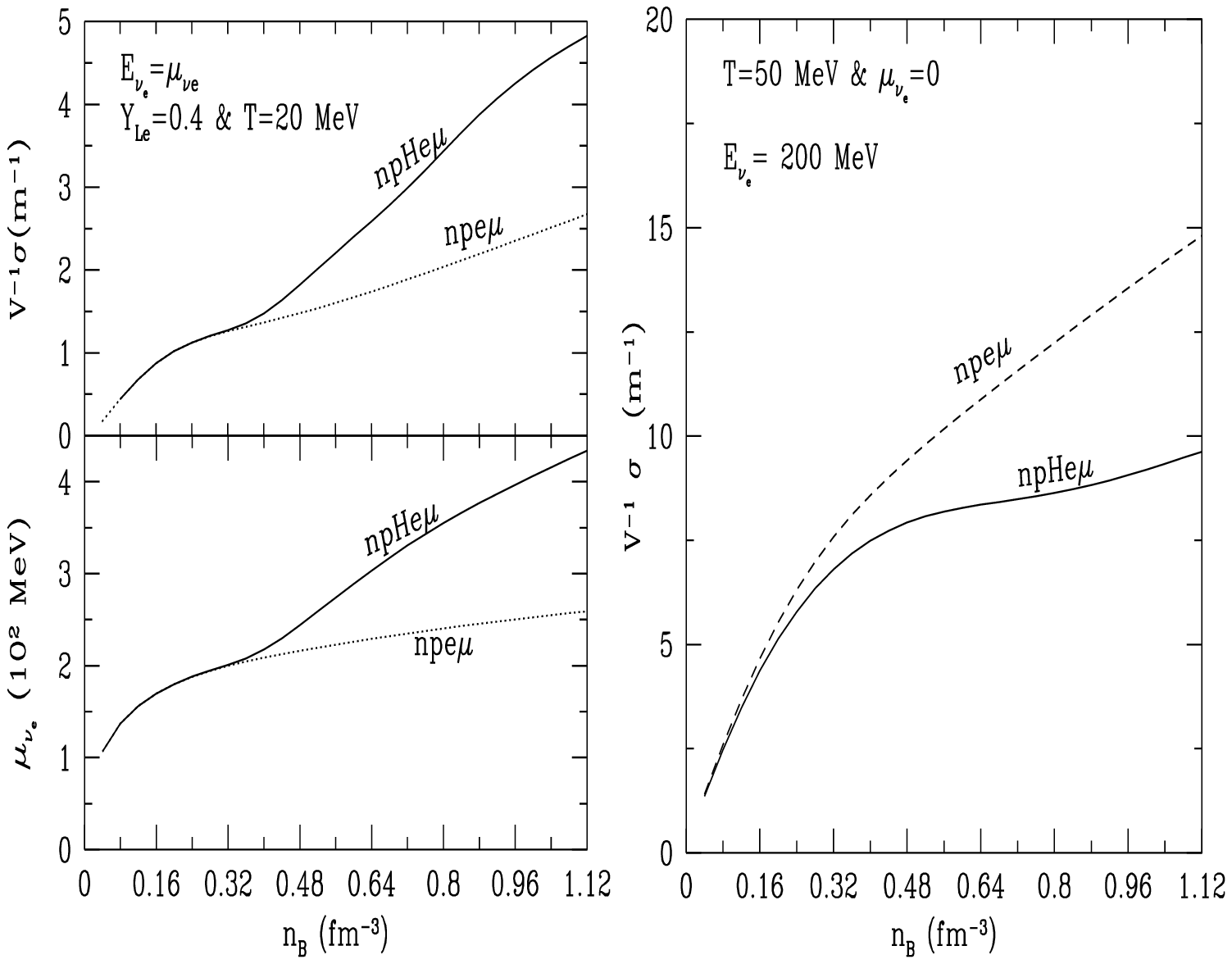}
\vskip -120pt
\begin{minipage}[t]{10cm}
\noindent \bf Fig.10.  \rm 
The upper left panel shows neutrino scattering cross sections in
neutrino-trapped matter appropriate for the deleptonization phase. The neutrino
energy is set equal to the local neutrino chemical potential, which is shown in
the lower left panel.  The right panel shows cross sections for thermal
neutrinos in neutrino-poor matter appropriate for the cooling phase. 
\end{minipage}
\vskip 4truemm

\begin{thebibliography}{99}\parindent=8truemm
\itemsep -1mm

\newcommand{\btem}{\bibitem}
\def\ap{{\it Ann. Phys.~}}
\def\apj{{\it Astrophys. J.~}}
\def\apjl{{\it Astrophys. J. Lett.~}}
\def\arns{{\it Ann. Rev. Nucl. Sci.~}}
\def\araa{{\it Ann. Rev. Astron. Astrophys.~}}
\def\apjs{{\it Astrophys. J. Suppl.~}}
\def\np{{\it Nucl. Phys.~}}
\def\pl{{\it Phys. Lett.~}}
\def\prp{{\it Phys. Rep.~}}
\def\prl{{\it Phys. Rev. Lett.~}}
\def\pr{{\it Phys. Rev.~}}
\def\nat{{\it Nature~}}
\def\aa{{\it Astron. \& Astrophys.~}}
\def\rmp{{\it Rev. Mod. Phys.~}}

\btem{prep} M. Prakash, I. Bombaci, Manju Prakash, P.J. Ellis, J.M. Lattimer 
and R. Knorren, nucl-th/9603042, {\it Phys. Rep.} (1996) in press.
\btem{comments} P.J. Ellis, J.M. Lattimer and M. Prakash, 
{\it Comments on Nucl. and Part. Phys.} (1996) in press.
\btem{mass} S.E. Thorsett, Z. Arzoumanian, M.M. McKinnon and J.H. Taylor, 
\apj {\bf 405} (1994) L29; 
M. H. van Kerkwijk, J. van Paradijs and E. J. Zuiderwijk,
\aa {\bf303} (1995) 497.
\btem{bww}
G.E. Brown, J.C. Weingartner and R.A.M.J. Wijers, \apj {\bf 463}
(1996) 297.
\btem{birth} A. Burrows and J.M. Lattimer,
\apj {\bf307} (1986) 178; 
A. Burrows, \arns {\bf 40} (1990) 181. 
\btem{LPPH} J.M. Lattimer, C.J. Pethick, M. Prakash, and P. Haensel, 
{\em Phys. Rev. Lett.}  {\bf 66} (1991) 2701. 
\btem{PPLP} Madappa Prakash, Manju Prakash, J.M. Lattimer and C.J. Pethick, 
\apjl {\bf 390} (1992) L77. 
\btem{Peth}
C.J. Pethick, \rmp {\bf 64} (1992) 1133.
\btem{Prak}
M. Prakash, \prp {\bf 242} (1994) 297. 
\btem{Ogel} H. \"Ogelman,  in 
{\it The Lives of the Neutron Stars}, 
M.A. Alpar, \"{U}. Kizilo\u{g}lu and J. van Paradijs (eds), 
(Dordrecht: Kluwer) 1995. 
\btem{HR}
J.P. Halpern and M. Ruderman, \apj {\bf 415} (1993) 286.
\btem{TolOv}
R.C. Tolman, {\it Proc. Nat. Acad. Sci. USA} {\bf 20} (1934) 3; 
J.R. Oppenheimer and G.M. Volkoff, \pr {\bf 55} (1939) 374. 
\btem{Wein} S. Weinberg, {\it Gravitation and Cosmology: Principles and
Applications of the General Theory of Relativity} (New York: Wiley) 1972.   
\bibitem{LPMY}
J.M. Lattimer, M. Prakash, D. Masak and A. Yahil, 
\apj {\bf 355} (1990) 241.
\bibitem{Kris} C. Kristian, {\it et al.}, \nat {\bf338} (1989) 234.
\bibitem{Pen} C. Pennypacker {\it et al.}, private communication (1990). 
\bibitem{Haen} 
P. Haensel, {\it Copernicus Astronomical Center Preprint} 
(1990). 
\bibitem{PRL} 
C.J. Pethick, D.G. Ravenhall and C.P. Lorenz, 
\np {\bf A584} (1995) 675.
%\btem{snth} A. Burrows and B. Fryxell, 
%\apjl {\bf 418} (1993) L33;
%J. Hayes and A. Burrows, 
%\apj {\bf 450} (1995) 830; 
%A. Burrows and J. Hayes, 
%\prl {\bf 76} (1995) 352; 
%M. Herant, W. Benz, J. Hix, C. Fryer and S.A. Colgate, 
%\apj {\bf 435} (1994) 339. 
\btem{bur} A. Burrows,
\apj {\bf 334} (1988) 891.
%\bibitem{rr} C.E. Rhoades and R. Ruffini, 
%\prl {\bf 32} (1974) 324.
\btem{bbw} G.E. Brown, S.W. Bruenn and J.C. Wheeler, {\it Comments Astrophys.}
{\bf16} (1992) 153.
\btem{tpl} V. Thorsson, M. Prakash and J.M. Lattimer, 
\np {\bf A572} (1994) 693. 
\btem{keiljan} W. Keil and H.T. Janka, 
\aa {\bf 296} (1994) 145. 
\btem{pcl} M. Prakash, J. Cooke and J.M. Lattimer, 
\pr {\bf D52} (1995) 661.
\btem{subside} N.K. Glendenning, 
\apj {\bf448} (1995) 797. 
\btem{bb94} G.E. Brown and H.A. Bethe, \apj {\bf 423} (1994) 659.
\btem{glenhyp} N.K. Glendenning, \np {\bf A493} (1989) 521;
J. Ellis, J.I. Kapusta and K.A. Olive,
\np {\bf B348} (1991) 345.
\btem{kapnel} D.B. Kaplan and A.E. Nelson, 
\pl {\bf B175} (1986) 57;
{\bf B179} (1986) 409(E).
\bibitem{gleqk} N.K. Glendenning, 
\pr {\bf D46} (1992) 1274. 
\btem{pcollapse} H.A. Bethe, G.E. Brown, J. Applegate and J.M. Lattimer,
\np {\bf A324} (1979) 487.
\btem{wff} R.B. Wiringa, V. Fiks and A. Fabrocini,
\pr {\bf C38} (1988) 1010.
\btem{MTI} T. Muto, T. Tatsumi and N. Iwamoto, Kyoto preprint 
KUNS-1382 (1996). 
\btem{exp87a} K. Hirata {\it et al.}, \prl {\bf58} (1987) 1490;
R.M. Bionta {\it et al.}, \prl {\bf58} (1987) 1494.
\btem{lcurve} S. Kumagai, T. Sigeyama, M. Hashimoto and K. Nomoto,
\aa {\bf 243} (1991) L13.
\btem{cc} K. Chen and S.A. Colgate, Los Alamos preprint 
LA-UR-95-2972 (1995).
\bibitem{thm} F.-K. Thielemann, M. Hashimoto and K. Nomoto,
\apj {\bf 349} (1990) 222.
\bibitem{bb} H.A. Bethe and G.E. Brown, 
\apj {\bf445} (1995) L129.
\btem{RP} S. Reddy and M. Prakash, 
\apj (1996) in press.
\bibitem{bkg} A. Burrows, D. Klein and R. Gandhi, 
\pr {\bf D45} (1992) 3361.  
\end{thebibliography}
\end{document}